\documentclass[prd,floatfix,showpacs, eqsecnum]{revtex4}
\usepackage{epsfig}

\newcommand{\eq}[1]{Eq.~(\ref{#1})}
\newcommand{\fig}[1]{Fig.~{\ref{#1}}}

\newcommand{\be}{\begin{equation}}
\newcommand{\ee}{\end{equation}}
\newcommand{\bea}{\begin{eqnarray}}
\newcommand{\eea}{\end{eqnarray}}
\newcommand{\ben}{\begin{eqnarray*}}
\newcommand{\een}{\end{eqnarray*}}

\newcommand{\DS}{Dyson-Schwinger }
\newcommand{\BS}{Bethe-Salpeter }
\newcommand{\ST}{Slavnov-Taylor }
\newcommand{\YM}{Yang-Mills }

\newcommand{\w}{\omega}
\newcommand{\e}{\varepsilon}
\newcommand{\al}{\alpha}
\newcommand{\ba}{\beta}
\newcommand{\ga}{\gamma}
\newcommand{\G}{\Gamma}
\newcommand{\de}{\delta}

\newcommand{\si}{\sigma}

\newcommand{\cs}{{\cal S}}
\renewcommand{\div}{\vec{\nabla}}

\newcommand{\s}[2]{{#1}\!\cdot\!{#2}}
\newcommand{\ov}[1]{\overline{#1}}
\newcommand{\dk}[1]
{\,\,\,\raisebox{-0.4ex}{\large $\bar{}$}\!\!d\,{#1}\,}

\begin{document}

\title{Three-quark confinement potential from the Faddeev equation}
\author{C.~Popovici}
\author{P.~Watson}
\author{H.~Reinhardt}
\affiliation{Institut f\"ur Theoretische Physik, Universit\"at
T\"ubingen, Auf der Morgenstelle 14, D-72076 T\"ubingen, Germany}
%\date{\today}
\begin{abstract}

In the heavy quark limit of Coulomb gauge QCD and by truncating the
Yang-Mills sector to include only dressed two-point functions, an 
analytic nonperturbative solution to the Faddeev equation for 
three-quark bound states in the case of equal quark separations is 
presented. A direct connection between the temporal gluon propagator
and the three-quark confinement potential is provided and it is shown
that only color singlet \emph {qqq} (baryon) states are physically 
allowed.
\end{abstract}
\pacs{11.10.St,12.38.Aw}
\maketitle

\section{Introduction}
\setcounter{equation}{0}

In order to understand the infrared properties of Quantum
Chromodynamics [QCD], the heavy quark sector is a useful area of 
study.  Among the heavy quark correlations, the most basic quantity 
is  the confinement potential between a quark-antiquark pair. 
At large separations, Wilson loops in lattice calculations exhibit an 
area law which corresponds to a linearly rising potential whose 
coefficient, the Wilsonian string tension, can be explicitly related
to a hadronic scale \cite{Sommer:1993ce}. Within continuous 
functional approaches in Coulomb gauge, recent investigations have
shown that (at least under truncation) there is a direct
connection between the Green's functions of \YM theory  and the
physical string tension that confines quarks \cite{Popovici:2010mb}.
In the Hamiltonian formalism, this relates to both the non-Abelian
color Coulomb potential \cite{Epple:2006hv}, and the temporal 
Wilson loop \cite{Pak:2009em}.

On the other hand, the potential that describes the interaction of 
three quarks  is much less studied then the potential between a 
quark-antiquark pair. Lattice simulations have indeed shown that
the \YM interaction gives rise to a linearly rising potential
between three quarks, but unfortunately a consensus regarding
the shape of the gluonic strings connecting the three quarks
is still missing: either the strings meet at the so-called Fermat 
point, which has minimal distances to the three sources
(so-called $Y$ configuration) 
\cite{Sommer:1985da,Bornyakov:2004uv,Takahashi:2002bw}, or
the $qqq$ potential is simply the sum of two-body interactions
(so-called $\Delta$ configuration) \cite{Cornwall:1996xr}. 
In the continuum, the only calculations of the three quark potential 
have been performed at perturbative level, within the so-called
potential non-relativistic QCD  approach \cite{Brambilla:1999xf}. 
In Ref. \cite{Brambilla:2009cd}, the authors considered the 
perturbative static potential of three heavy quarks and found that 
up to NLO, the potential is just the sum of the two-body 
contributions, whereas at NNLO the three-body contributions do
appear, signaling the importance of three-body interactions for 
understanding the shape of the string in the infrared regime.     

The Faddeev equation \cite{Faddeev:1960su} and its subsequent
developments \cite{Taylor:1966zza,Boehm:1976ya}
(for an extended review see \cite{Loring:2001kv})
provide a  general formulation of the relativistic three-body 
problem. It is a bound state equation (the direct analogue of the
homogeneous two-body Bethe-Salpeter equation) and it has been
efficiently applied in QCD to study baryon states, via the Green's 
functions of the theory. Typically, these studies are performed in
Landau gauge and, due to the complexity of the equations, 
they have been mainly restricted to rainbow-ladder truncation, 
where the kernel is reduced to the single exchange of a dressed
gluon. Within this approximation and by employing 
phenomenological Ans\"atze for the \YM part of the theory,
 the nucleon and $\Delta$ properties have been studied 
\cite{Hellstern:1997pg,Oettel:2000jj,Eichmann:2009qa,
Nicmorus:2010mc}. Other simplifications include the three-body 
spectator formalism \cite{Loring:2001kv}, a Salpeter-type equation 
with instantaneous interaction \cite{Stadler:1997iu} or the diquark 
correlations \cite{Anselmino:1992vg}. Already from the \BS studies
for mesons, it is known that truncating the kernel is not a simple
task -- the truncation has to be consistent with the symmetry 
properties of the theory, e.g., the axialvector Ward-Takahashi 
identity must be satisfied. In contemporary studies, the \BS kernel
has been considered  beyond the rainbow-ladder truncation
for meson states, including both vertex corrections \cite
{Watson:2004kd,Williams:2009wx,Bender:2002as,Bhagwat:2004hn,
Matevosyan:2006bk,Bender:1996bb} as well as unquenching effects
\cite{Williams:2009wx,Fischer:2005en,Watson:2004jq,Fischer:2009jm}
and it was found that (apart from meson decay induced by 
unquenching \cite{Watson:2004jq}) the rainbow-ladder approximation
works surprisingly well.

For many years, Coulomb gauge studies have been recognized as a
promising avenue with which to investigate the nonperturbative
regime of QCD \cite{Abers:1973qs}. In this gauge, the 
Gribov-Zwanziger scenario of confinement becomes particularly 
relevant \cite{Gribov:1977wm,Zwanziger:1995cv,Zwanziger:1998ez}. 
In this picture, the temporal component of the gluon propagator 
becomes infrared enhanced, providing for a long range confining force,
while the transversal spatial component is infrared suppressed, thus 
explaining the absence of the asymptotic states in the spectrum.  
Coulomb gauge is physical, in the sense that in this gauge the system 
reduces naturally to the physical degrees of freedom (explicitly
demonstrated in \cite{Zwanziger:1998ez}). Moreover, within the first
order functional formalism it has been shown that the total charge 
of the system is conserved and vanishing, and the well-known energy
divergence problem disappears \cite{Reinhardt:2008pr}.  Within this
approach  the \DS equations for the \YM part of the theory have been
derived \cite{Watson:2006yq,Watson:2007vc}, together with the \ST
identities \cite{Watson:2008fb} and perturbative results have been 
provided \cite{Watson:2007mz}. In addition, the quark sector has 
been also investigated, within perturbation theory
\cite{Popovici:2008ty} as well as in the heavy mass limit 
\cite{Popovici:2010mb}. On the lattice, initial calculations for the
\YM propagators have also become available 
\cite{Burgio:2008jr,Quandt:2008zj} (see also 
\cite{Nakagawa:2009zf,Cucchieri:2000gu,Langfeld:2004qs}). In
particular, the results indicate that the temporal component of the
gluon propagator is largely independent of energy (due to 
noncovariance, in Coulomb gauge the propagators are in general 
dependent on both the energy and spatial momentum), and it is 
consistent with a $1/\vec q^4$ behavior in the infrared. Moreover,
the spatial equal-time gluon propagator is found to be vanishing
in the infrared, in agreement with the Gribov's formula
\cite{Zwanziger:1991gz, Burgio:2008jr}. The lattice calculations
support the results obtained from the variational method to the
Hamiltonian approach in \YM theory \cite{Feuchter:2004mk,
Epple:2006hv,Epple:2007ut,Szczepaniak:2001rg}.

In this  paper, we extend a previous investigation of the heavy 
quark system in Coulomb gauge \cite{Popovici:2010mb}.
There, the \BS equation for $\bar qq$ bound states was studied
with a heavy quark mass expansion  (which underlines the Heavy
Quark Effective Theory [HQET] 
\cite{Neubert:1993mb,Mannel:1992fx,Grinstein:1991ap})
at leading order, and  a direct connection between the temporal 
gluon propagator and the string tension was found. Following the
same approach, we consider in this work the Faddeev equation for
three-quark systems  in Coulomb gauge, in the symmetric case
(i.e.,  equal spatial separations between quarks) and with the
inclusion of only two-body interactions, at leading order in the 
mass expansion. We will use the results inspired by the lattice 
for the Green's functions of the \YM sector  and in addition, 
we will employ our previous findings, in particular that 
nonperturbatively the temporal quark-gluon vertex remains bare
under truncation and the kernel of the \BS equation reduces to the
ladder approximation. In this truncated system, we will provide an
exact solution to the Faddeev equation, from which the confining
potential between three quarks emerges, and we will show that
\emph{qqq} bound states can only exist for $N_c=3$ colors, i.e.
color singlet baryons. (In \cite{Popovici:2010mb} it was shown
that only color-singlet meson and $SU(2)$ $qq$ ``baryon'' states
have finite energy.)

The organization of this paper is as follows. In Sec.~II we briefly
review relevant results for the heavy quark systems. Starting
with the generating functional of Coulomb gauge QCD at leading
order in the mass expansion, we review the main steps in the 
derivation of the heavy quark propagator and the temporal
quark-gluon vertex needed in this work. In Sec.~III the Faddeev 
equation for three-quark states  is considered. In addition to
solving the equation, the pole structure of the quark-baryon vertex
is analyzed and an interesting similarity with the $\bar qq$ system
is discussed. Moreover, a direct connection between the temporal 
gluon propagator and the physical string tension is found. In 
Sec.~IV a short summary and the concluding remarks will be 
presented. Some technical details are given in the Appendix.

\section{Heavy quark mass expansion}

In this section, let us briefly review some relevant results from 
\cite{Popovici:2010mb}. The notations and conventions used in this
work are those established in Refs.~\cite{Popovici:2010mb,
Popovici:2008ty, Watson:2008fb,Watson:2007vc,Watson:2006yq}. 
We work in Minkowski space, with the metric 
$g_{\mu\nu}=\mbox{diag}(1,-\vec{1})$. Roman letters 
($i$, $j$,\ldots) refer to spatial indices and superscripts
($a$, $b$,\ldots) stand for color indices in the adjoint  
representation of the gauge group.  Unless otherwise specified,
the Dirac spinor, flavor and (fundamental) color indices are denoted
with a common index ($\al, \ba\dots$). Configuration space 
coordinates may be denoted with subscript ($x$, $y$,\ldots) when no
confusion arises.   The Dirac $\ga$-matrices satisfy the Clifford 
algebra $\{\ga^\mu,\ga^\nu\}=2g^{\mu\nu}$. The notation $\ga^{i}$ 
refers to the spatial component, where the minus sign arising from
the metric has been explicitly extracted when appropriate. 

The explicit quark contribution to the full QCD generating 
functional within our conventions, can be written 
\cite{Popovici:2008ty}
\bea
Z[\ov{\chi},\chi]&=&\int{\cal D}\Phi\exp{\left\{\imath
\int d^4x \ov{q}_\al(x)\left[\imath\ga^0D_0
+\imath\s{\vec{\ga}}{\vec{D}}-m\right]_{\al\ba}q_\ba(x)
\right\}}
\nonumber\\&&\times
\exp{\left\{\imath\int d^4x\left[\ov{\chi}_\al(x)q_\al(x)+
\ov{q}_\al(x)\chi_\al(x)\right]+\imath{\cal S}_{YM}\right\}}.
\label{eq:genfunc}
\eea
In the above, ${\cal D}\Phi$ generically denotes the functional 
integration measure over all fields and $\cs_{YM}$ is the \YM 
contribution to the generating functional.  $q_\al$ denotes the
full quark field, $\bar q_\al$ is the conjugate (or antiquark) 
field, and  $\chi_\al$, $\ov{\chi}_\al$ are the corresponding 
sources. The temporal and spatial components of the covariant 
derivative (in the fundamental color representation) are given by
\bea
D_0&=&\partial_{0}-\imath gT^a\si^a(x),\nonumber\\
\vec{D}&=&\div+\imath gT^a\vec{A}^a(x),
\eea
where $\vec{A}$ and $\si$ refer to the spatial and temporal
components of the gluon field, respectively.  $f^{abc}$ are the 
structure constants of the $SU(N_c)$ group, with the Hermitian
generators $T^a$, satisfying $[T^a,T^b]=\imath f^{abc}T^c$ and
normalized via $\mbox{Tr}(T^aT^b)=\de^{ab}/2$. For later use we
introduce the Casimir factor associated with the quark gap equation:
\be
C_F=\frac{N_c^2-1}{2N_c}.
\label{eq:casimir1}
\ee

In the following, we briefly sketch the derivation of the quark 
propagator, in the heavy mass limit. For an extended discussion,
the reader is referred to the original work \cite{Popovici:2010mb}.
We start by performing the following decomposition of the quark
field
\bea
q_\al(x)=e^{-\imath mx_0}\left[h(x)+H(x)\right]_\al,
&h_\al(x)=e^{\imath mx_0}\left[P_+q(x)\right]_\al,
&H_\al(x)=e^{\imath mx_0}
\left[P_-q(x)\right]_\al
\label{eq:qdecomp}
\eea
(similarly for the antiquark field) and introduce the two components
$h$ and $H$, with the help of the spinor projectors
\be
P_\pm=\frac{1}{2}(\openone\pm\ga^0).
\ee
This corresponds to a particular case of the heavy quark transform
underlying HQET with the velocity vector $v^{\mu}=(1,0,0,0)$
\cite{Neubert:1993mb}, but in the functional approach here it can 
be regarded simply as an arbitrary decomposition that will turn out 
to be very useful in Coulomb gauge (precisely, this will lead to the
suppression of the spatial gluon propagator in the mass 
expansion, see below).

We now insert the  decomposition, \eq{eq:qdecomp}, into the 
generating functional \eq{eq:genfunc}, integrate out the 
$H$-fields, and make an expansion in the heavy quark mass 
(throughout this work,  we will use the standard term  
``mass expansion'', instead of
``expansion in the \emph{inverse} mass''). At leading order, 
we get the following expression:
\bea
Z[\ov{\chi},\chi]&=&\int{\cal D}\Phi\exp{\left\{\imath
\int d^4x\ov{h}_\al(x)
\left[\imath\partial_{0x}+gT^a\si^a(x)\right]_{\al\ba}
h_\ba(x)\right\}}
\nonumber\\&&\times
\exp{\left\{\imath
\int d^4x\left[e^{-\imath mx_0}\ov{\chi}_\al(x) h_\al(x)
+e^{\imath mx_0}\ov{h}_\al(x)\chi_\al(x)\right]+
\imath{\cal S}_{YM}\right\}}+{\cal O}\left(1/m\right),
\label{eq:genfunc4}
\eea
where  the temporal component of the covariant derivative $D_0$ has
been written explicitly. In the above, we have kept the full quark 
and antiquark sources (rather than  the ones corresponding to the
components of the quark field, introduced in HQET). This means that
we can use the full  gap, \BS and  Faddeev equations of QCD but
replace the kernels, propagators and vertices and restrict to the
leading order in the mass expansion.

In full QCD (i.e., Coulomb gauge within second order formalism, 
without the mass expansion and derived from the first order 
formalism results of Ref.~\cite {Popovici:2008ty}),  the quark gap
equation  is given by [$\dk{\w}=d^4\w/(2\pi)^4$]: 
\bea
\G_{\ov{q}q\al\de}(k)&=\G_{\ov{q}q\al\de}^{(0)}(k)+\int\dk{\w}&
\left\{\G_{\ov{q}q\si\al\ba}^{(0)a}(k,-\w,\w-k)W_{\ov{q}q\ba\ga}
(\w)\G_{\ov{q}q\si\ga\de}^{b}(\w,-k,k-\w)W_{\si\si}^{ab}(k-\w)
\right.\nonumber\\&&\left.
+\G_{\ov{q}qA\al\ba i}^{(0)a}(k,-\w,\w-k)W_{\ov{q}q\ba\ga}(\w)
\G_{\ov{q}qA\ga\de j}^{b}(\w,-k,k-\w)W_{AAij}^{ab}(k-\w)
\right\}
\label{eq:gap}
\eea
($W_{AA}$ is the spatial gluon propagator, which will not 
be regarded  here). The gap equation is supplemented by the \ST
identity, which follows from the invariance of the action under a
Gauss-BRST transform \cite{Popovici:2010mb}. In Coulomb gauge,
this identity reads:
\bea
k_3^0\G_{\ov{q}q\si\al\ba}^{d}(k_1,k_2,k_3)&=&
\imath\frac{k_{3i}}
{\vec{k}_3^2}\G_{\ov{q}qA\al\ba i}^{a}(k_1,k_2,k_3)
\G_{\ov{c}c}^{ad}(-k_3)\nonumber\\&&
+\G_{\ov{q}q\al\de}(k_1)\left[\tilde{\G}_{\ov{q};\ov{c}cq}^{d}
(k_1+q_0,k_3-q_0;k_2)+\imath gT^d\right]_{\de\ba}
\nonumber\\&&
+\left[\tilde{\G}_{q;\ov{c}c\ov{q}}^{d}(k_2+q_0,k_3-q_0;k_1)-
\imath gT^d\right]_{\al\de}\G_{\ov{q}q\de\ba}(-k_2)
\label{eq:stid}
\eea
where $k_1+k_2+k_3=0$, $q_0$ is an arbitrary energy injection 
scale (arising from the noncovariance of Coulomb gauge
\cite{Watson:2008fb}), $\G_{\ov{c}c}$ is the ghost proper two-point
function,  $\tilde{\G}_{\ov{q};\ov{c}cq}$ and
$\tilde{\G}_{q;\ov{c}c\ov{q}}$ are ghost-quark kernels associated 
with the Gauss-BRST transform. 

Now, as a consequence of the Coulomb gauge decomposition,
\eq{eq:qdecomp}, the part of the generating functional given by
\eq{eq:genfunc4} corresponding to the tree-level spatial quark 
gluon vertex $\G_{\ov{q}qA}^{(0)}$ is contained within the $O(1/m)$
contribution which is here neglected. Under the further assumption
that the pure \YM vertices may be neglected, the \DS equation for
the nonperturbative spatial quark-gluon vertex then furnishes the 
result that $\G_{\bar qqA}\sim O(1/m)$ (see \cite{Popovici:2010mb} 
for a complete discussion and justification of this truncation). 
Similarly, the ghost-quark kernels can be neglected. Thus, under 
our truncation scheme, the \ST identity reduces to
\be
k_3^0\G_{\ov{q}q\si\al\ba}^{d}(k_1,k_2,k_3)=
\G_{\ov{q}q\al\de}(k_1)\left[\imath gT^d\right]_{\de\ba}-
\left[\imath gT^d\right]_{\al\de}\G_{\ov{q}q\de\ba}(-k_2)
+{\cal O}\left(1/m\right).
\ee
This is then inserted into \eq{eq:gap}, together with the 
tree-level quark  proper two-point function
\be
\G_{\ov{q}q\al \ba}^{(0)}(k)=\imath\de_{\al\ba}\left[k_0-m\right]
+{\cal O}\left(1/m\right)
\label{eq:gaptree}
\ee
and the tree level quark gluon vertex 
\be
\G_{\ov{q}q\si\al\ba}^{(0)a}(k_1,k_2,k_3)=
\left[gT^a\right]_{\al\ba}+{\cal O}\left(1/m\right)
\label{eq:feyn0}
\ee
that follow from the generating functional \eq{eq:genfunc4}. The 
general form of the nonperturbative temporal gluon
propagator is given by \cite{Watson:2007vc}:
\be
W_{\si\si}^{ab}(\vec k)=
\de^{ab}\frac{\imath}{\vec{k}^2}D_{\si\si}(\vec{k}^2).
\label{eq:Wsisi}
\ee
Lattice results, and also more formal consideration in continuum
show that  the  dressing function  $D_{\si\si}$ has some part that
is independent of energy \cite{Cucchieri:2000hv} and moreover,
$D_{\si\si}$ is  infrared divergent and likely to behave as 
$1/\vec{k}^2$ for vanishing $\vec{k}^2$ (the explicit form of
$D_{\si\si}$ will only be needed in the last step of the 
calculation). Putting  all this together, we find the following
solution to \eq{eq:gap} for the heavy quark propagator:
\bea
W_{\ov{q}q\al\ba}(k_0)=\frac{-\imath\de_{\al\ba}}{\left[k_0-m-
{\cal I}_r+\imath\e\right]}+{\cal O}\left(1/m\right),
\label{eq:quarkpropnonpert}
\eea
with the (implicitly regularized, denoted by ``$r$'') constant
[$\dk{\vec{\w}}=d^3\vec{\w}/(2\pi)^3$]
\be
 {\cal I}_r =\frac{1}{2}g^2C_F
\int_r\frac{\dk{\vec{\w}}D_{\si\si}(\vec{\w})}{\vec{\w}^2}
+{\cal O}\left(1/m\right).
\label{eq:iregularized}
\ee
When solving \eq{eq:gap}, the ordering of the integration  is set
such that the temporal integral is performed first, under the 
condition that the spatial integral is regularized and finite. 
Inserting the solution \eq{eq:quarkpropnonpert} into the \ST
identity, we find that the temporal quark-gluon vertex remains
nonperturbatively bare:
\be
\G_{\ov{q}q\si\al\ba}^{a}(k_1,k_2,k_3)=
\left[gT^a\right]_{\al\ba}+{\cal O}\left(1/m\right).
\label{eq:feyn}
\ee

The propagator \eq{eq:quarkpropnonpert} has a couple of striking
features, which have been emphasized in \cite{Popovici:2010mb}.
Firstly, due to the mass expansion, we only have a single pole in
the complex $k_0$-plane, as opposed to the conventional quark
propagator, which possesses a pair of simple poles. Hence, it is
necessary to explicitly define the Feynman prescription. From
\eq{eq:quarkpropnonpert} it then follows that the closed quark
loops (virtual quark-antiquark pairs) vanish due  to the energy 
integration, which implies that the theory is quenched in the 
heavy mass limit:
\be
\int\frac{dk_0}{\left[k_0-m- {\cal I}_r+\imath\e\right]
\left[k_0+p_0-m- {\cal I}_r+\imath\e\right]}=0.
\label{eq:tempint}
\ee
Secondly, the propagator \eq{eq:quarkpropnonpert} is diagonal in the
outer product of the fundamental color, flavor and spinor spaces --
physically this corresponds to the decoupling of the spin from the
heavy quark system. In fact,  $W_{\ov{q}q}^{(0)}$ is identical to the
heavy quark tree-level propagator \cite{Neubert:1993mb} up to
the appearance of the mass term, and this is due to the fact that
in HQET one uses the sources for the large $h$-fields directly, 
while we retain the sources of the full quark fields. Finally, let
us emphasize that the position of the  pole has no physical meaning
since the quark can never be on-shell. The poles in the quark 
propagator are situated at infinity (thanks  to ${\cal I}_r$ as the
regularization is removed) meaning that either one requires infinite
energy to create a quark from the vacuum or, if a a hadronic system
is considered, only the relative energy is important. Indeed, it was
shown some time ago \cite{Adler:1984ri} that the divergence of the
absolute energy has no physical meaning and only the relative energy 
(derived from the \BS equation) must be considered. It is precisely
the cancellation of these divergent constants that distinguishes
between physical and unphysical poles.

We also note that for the antiquark propagator  the opposite 
Feynman prescription is assigned such that the \BS equation for the
quark-antiquark states has a physical interpretation of a bound
state equation. There, the quark and the antiquark do not create a
virtual quark-antiquark pair, but a system composed of two separate
unphysical particles (in the sense that they are not connected by a
primitive vertex). Moreover, the \BS kernel reduces to the ladder
truncation \cite{Popovici:2010mb}. The reason is the cancellation of
the so-called crossed box contributions (i.e., nonplanar diagrams
that contain any combinations of nontrivial interactions allowed
within our truncation scheme) due to the temporal integration
performed over multiple propagators with the same relative sign for
the Feynman prescription (similar to \eq{eq:tempint}, but in this
case the terms  originate from internal quark or antiquark 
propagators).

\section{Faddeev equations for three-quark states}

\begin{figure}[t]
\vspace{0.5cm}
\includegraphics[width=1.00\linewidth]{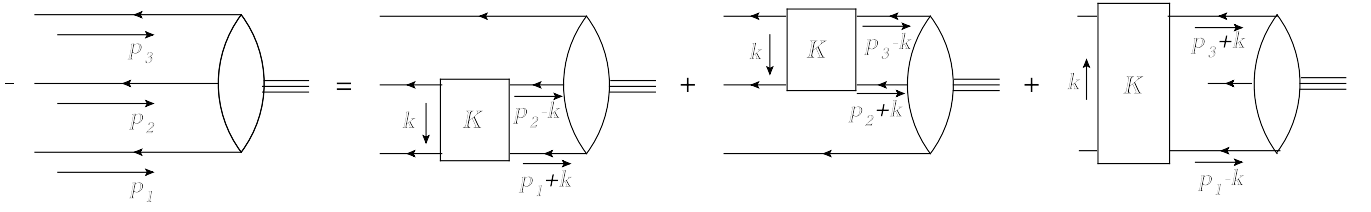}
\caption{\label{fig:faddeev}
Faddeev equation for three quark bound states. Solid lines represent
the quark propagator, the box represents the diquark kernel $K$ and
the ellipse represents the Faddeev vertex function with the bound
state leg depicted by a triple-line. See text for details.}
\end{figure}

Let us now consider the Faddeev equation for three-quark bound 
states. In this work, we employ only the permuted two quark kernels
$K$ (which coincide with kernel appearing in the \BS equation for
diquark states) and neglect the three-quark irreducible diagrams, 
i.e., genuine three-body forces. This approximation is  also 
motivated by the fact that  in the quark-diquark model the binding 
energy is assumed to be mainly provided by the two-quark correlations
\cite{Reinhardt:1989rw}. In this truncation, the Faddeev equation 
reads (see also \fig{fig:faddeev}):
\bea
\G_{\al\ba\ga}(p_1,p_2,p_3;P)&=&
-\int\dk{k}
\left\{K_{\ba\al;\al^{'}\ba^{'}}(k)
W_{\bar qq\al^{'}\al^{''}}(p_1+k)W_{\bar qq\ba^{'}\ba^{''}}(p_2-k)
\G_{\al^{''}\ba^{''}\ga}(p_1+k,p_2-k,p_3;P)\right.\nonumber\\
&+&K_{\ga\ba;\ba^{'}\ga^{'}}(k)
W_{\bar qq\ba^{'}\ba^{''}}(p_2+k)W_{\bar qq\ga^{'}\ga^{''}}(p_3-k)
\G_{\al\ba^{''}\ga^{''}}(p_1,p_2+k,p_3-k;P)\nonumber\\
&+&\left. K_{\al\ga;\ga^{'}\al^{'}}(k)
W_{\bar qq\ga^{'}\ga^{''}}(p_3+k)W_{\bar qq\al^{'}\al^{''}}(p_1-k)
\G_{\al\ba^{''}\ga^{''}}(p_1-k,p_2,p_3+k;P)
\right\}
\label{eq:faddeev.gen}
\eea
where $p_1,p_2,p_3$ are the momenta of the quarks, $P=p_1+p_2+p_3$
is the pole 4-momentum of the bound baryon state and $\G$ is the 
so-called quark-baryon Faddeev vertex for the particular bound state
under consideration and whose indices denote explicitly only its
quark content. Due to the fact that in the heavy mass limit the
spin degrees of freedom decouple from the system, at leading order
in the mass expansion the Faddeev baryon amplitude $\G_{\al\ba\ga}$
becomes a Dirac scalar, similar to the heavy quark propagator
\eq{eq:quarkpropnonpert}. The explicit momentum dependence of the
kernels $K$ is abbreviated for notational convenience. As in the
homogeneous \BS equation, the integral equation depends only
parametrically on the total four momentum $P$.

As discussed at the end of the previous Section,  the kernel $K$
reduces to the ladder approximation (constructed via  gluon 
exchange) and it reads
\be
K_{\ba\al;\al^{'}\ba^{'}}(k)= 
\G_{\bar qq\si \al\al^{'}}^{a}W_{\si\si}^{ab}(\vec k) 
\G_{\bar qq\si\ba\ba^{'}}^{b}=g^2
T_{\al\al^{'}}^{a}W_{\si\si}^{ab}(\vec k) T_{\ba\ba^{'}}^{b}
\label{eq:kernel}
\ee
with the temporal gluon propagator and the temporal quark-gluon
vertex given by \eq{eq:Wsisi} and \eq{eq:feyn}, respectively. Similar
to the \BS equation for meson bound states, the energy independence
of this propagator will turn to be crucial in the derivation of the
confining potential.

Let us now have a closer look at the energy dependence of the
equation \eq{eq:faddeev.gen}. In the meson case, since the \BS
kernel was energy independent, it was straightforward to show that
the \BS vertex itself did not contain an energy dependent part. This
observation was then  used to calculate the confining potential from
the $\bar qq$ \BS equation, via a simple analytical integration over
the relative energy variable. Unfortunately this approach cannot be
extended to baryon states: despite the  instantaneous kernel, a
relative energy dependence still remains and thus one cannot assume
an energy-independent Faddeev vertex. Therefore, in order to proceed,
we make the following separable Ansatz for the Faddeev vertex:
\be
\G_{\al\ba\ga}( p_1, p_2,  p_3;P)=\Psi_{\al\ba\ga}
\G_{t}( p_1^0, p_2^0, p_3^0;P)
\G_{s}( \vec p_1,\vec  p_2, \vec  p_3;P)
\label{eq:qqqbaryon}
\ee
where we have introduced a purely antisymmetric (in the quark legs)
color factor  $\Psi$  (the possible baryon color index is omitted) 
and the symmetric (Dirac scalar) temporal and spatial components 
$\G_t$ and $\G_s$, respectively.

Inserting the explicit form of the kernel \eq{eq:kernel} and the
quark-baryon vertex Ansatz \eq{eq:qqqbaryon}, the Faddeev equation
\eq{eq:faddeev.gen} can be explicitly written as (for simplicity we
drop the label $P$ in the arguments of the vertex functions):
\bea
\G_{\al\ba\ga}(p_1,p_2,p_3)&=&\!-g^2
T_{\al\tau}^{a} T_{\ba\kappa}^{a}\Psi_{\tau\kappa\ga}
\!\int\!\!\dk{k}
W_{\si\si}(\vec k)
W_{\bar qq}(p_1+k)W_{\bar qq}(p_2-k)
\G_{t}( p_1^0+k_0, p_2^0-k_0, p_3^0)
\G_{s}( \vec p_1+\vec k,\vec p_2-\vec k, \vec  p_3)
\nonumber\\
&+&\textrm{~cyclic~permutations},
\label{eq:faddeev.gen1}
\eea
where the explicit color structure has been extracted 
($W_{\si\si}^{ab}=\de^{ab}W_{\si\si}, 
W_{\bar qq \al\ba}=\de_{\al\ba} W_{\bar qq}$).

With the help of the Fierz identity for the generators $T^a$
\be
2\left[T^a\right]_{\al\ba}\left[T^a\right]_{\de\ga}=\de_{\al\ga}
\de_{\de\ba}-\frac{1}{N_c}\de_{\al\ba}\de_{\de\ga},
\label{eq:fierz}
\ee 
the color structure can be written as
\be
T_{\al\al^{'}}^{a} T_{\ba\ba^{'}}^{a}\Psi_{\al^{'}\ba^{'}\ga}=
-C_{B} \Psi_{\al\ba\ga}
\ee
with $C_B=(N_c+1)/2N_c$, where $N_c$ is the number of colors, yet to
be identified (i.e., the baryon is not assumed to be a color singlet).

In the next step we perform the Fourier transform for the spatial 
part of the equation, recalling that the heavy quark propagator is
only a function of energy. We define the coordinate space vertex
function via its Fourier transform
\be
\G_{s}(\vec{p_1},\vec{p_2},\vec{p_3})=
\int d\vec x_1d \vec x_2  d\vec x_3
e^{-\imath\vec p_1\cdot\vec x_1-\imath\vec p_2\cdot\vec x_2
-\imath\vec p_3\cdot\vec x_3}
\G_{s}(\vec{x}_1,\vec{x}_2,\vec{x}_3)
\ee
(similarly for $W_{\si\si}$, as in \cite{Popovici:2010mb}) such that 
\bea
\int\dk{\vec k} W_{\si\si}(\vec k)
\G_{s}( \vec p_1+\vec k,\vec p_2-\vec k, \vec  p_3)=
\int \dk{\vec x_1}\dk{\vec x_2} \dk{\vec x_3}
e^{-\imath\vec p_1\cdot\vec x_1-\imath\vec p_2\cdot\vec x_2
-\imath\vec p_3\cdot\vec x_3}
W_{\si\si}(\vec x_2-\vec x_1)\G_{s}(\vec{x}_1,\vec{x}_2,\vec{x}_3).
\eea
Clearly, the component $\G_s$ trivially simplifies (as before, we
have separated the temporal and spatial integrals, under the
assumption the spatial integral is regularized and finite) and the
equation \eq{eq:faddeev.gen1} reduces to [$\dk{k_0}=d k_{0}/(2\pi)$]:
\bea
\G_{t}(p_1^0,p_2^0,p_3^0)&=&g^2C_B W_{\si\si}(\vec x_2-\vec x_1)
\int\dk{k_0}
W_{\bar qq}(p_1^0+k_0)W_{\bar qq}(p_2^0-k_0)
\G_{t}(p_1^0+k_0,p_2^0-k_0,p_3^0) 
+\textrm{~cyclic~permutations}.
\nonumber\\
\label{eq:faddeev.coordinate}
\eea
At this point we make a further simplification, motivated by the
symmetry of the three-quark system: we restrict to a particular
geometry, namely to equal quark separations, i.e. 
$|\vec r|= |\vec x_2-\vec x_1|=| \vec x_3-\vec x_2| 
=| \vec x_1-\vec x_3|$.  By inserting the explicit form of the
quark propagators, \eq{eq:quarkpropnonpert}, we have
\bea
&&\G_{t}(p_1^0,p_2^0,p_3^0)= -g^2C_B W_{\si\si}(|\vec r|)
\int\dk{k_0}
\frac{\G_{t}(p_1^0+k_0,p_2^0-k_0,p_3^0)}
{\left[p_1^0+k_0-m-{\cal I}_r+\imath\e\right]
\left[p_2^0-k_0-m-{\cal I}_r+\imath\e\right]}
+\textrm{~cyclic~permutations}.
\nonumber\\
\label{eq:faddeev.simpl}
\eea
Assuming that the vertex $\G_t$  is symmetric under permutation of
quark legs, an Ansatz  that satisfies this equation is:
\be
\G_{t}(p_1^0,p_2^0,p_3^0)= \sum_{i=1,2,3}
\frac{1}{2P_0-3(p_i^0 +m+{\cal I}_r)+\imath\e}.
\label{eq:gammat}
\ee
Since the explicit derivation is rather technical, we only give here
the solution and defer the details to the Appendix. 

Notice that in the expression \eq{eq:gammat} there are simple poles
(in the energy) present. These poles however do not occur for finite
energies and cannot be physical. As discussed, this is also the case
for the quark propagator. Intuitively, when a single heavy quark is
pulled apart from the system, the $qqq$ state becomes equivalent
(i.e.,  it has the same color quantum numbers) to the $\bar qq$
system in the sense that the remaining two quarks form a diquark 
which for $N_c=3$ would be a color antitriplet configuration, and
hence the physical interpretation of the vertex \eq{eq:gammat} can be
directly related to the heavy quark propagator
\eq{eq:quarkpropnonpert}: the presence of the single pole in
\eq{eq:gammat} simply means that this cannot have the meaning of
physical propagation (this would require a covariant double pole).
Moreover, the divergent constant ${\cal I}_r$ appearing in the
absolute energy does not contradict the physics -- the only relevant
quantity is the relative energy of the three quark system.

With this Ansatz at hand, we  return to  the formula
\eq{eq:faddeev.simpl}, insert the definitions \eq{eq:Wsisi} and
\eq{eq:iregularized} for $W_{\si\si}(\vec x)$ and  ${\cal I}_r$, and
arrive at the  following solution for the bound state energy $P_0$,
in the case of equal quark separations:
\be
P_0=3
m+\frac{3}{2}g^2\int\dk{\vec\w}\frac{D_{\si\si}(\vec\w)}{\vec\w^2}
\left[C_F- 2 C_B e^{\imath\vec\w\cdot \vec r}\right].
\label{eq:P0solution}
\ee
Since the quarks cannot be prepared as isolated states, the only
possibilities for the $qqq$ state are either that the system is
confined (i.e.,  the bound state energy $P_0$ increases with the
separation), or the system is  physically not allowed (i.e.,  the 
energy $P_0$ is infinite). From the formula \eq{eq:P0solution} and
knowing that $D_{\si\si}(\vec\w)$ is infrared enhanced , it is clear
that in order to have an infrared confining solution (corresponding
to a convergent three-momentum integral), the condition
\be
C_B=\frac{C_F}{2}
\ee
must be satisfied. This is fulfilled for $N_c=3$ colors, implying
that $\Psi_{\al\ba\ga}=\e_{\al\ba\ga}$ and that the baryon is a color
singlet (confined) bound state of three quarks; otherwise, for
$N_c\ne 3$ the energy of the  the system is infinite for any
separation $|\vec r|$.

Assuming that in the infrared $D_{\si\si}(\vec\w)= X/\vec \w^2$
(as indicated by the lattice data \cite{Burgio:2008jr,Quandt:2008zj,
Nakagawa:2009zf,Cucchieri:2000gu,Langfeld:2004qs}
and by the variational calculations in the continuum
\cite{Epple:2006hv}), where $X$ is some combination of constants, 
it is straightforward to perform the integration on the right hand
side of \eq{eq:P0solution}, with the result that for large separation
$|\vec r|$:
\be
P_0=3m+\frac{3}{2}\frac{g^2C_F X}{8\pi}|\vec r|.
\label{eq:bound_state}
\ee
This mimics the previous findings for $\bar qq$ systems, namely that
there exists a direct connection between the string tension and the
nonperturbative \YM Green's functions (at least under truncation). In
this case, the standard term ``string tension'' refers to the
coefficient of the three-body linear confinement term 
$\si_{3q}|\vec r|$. Also, comparing with the result of 
Ref. \cite{Popovici:2010mb}, we find that the string tension
corresponding to the $qqq$ system is $3/2$ times that of the
$\bar qq$. To our knowledge, no direct comparison between the string
tensions  of the two systems has been made and hence this relation
would be interesting to investigate on the lattice. The appearance of
three times the quark mass stems from the presence of the mass term
in the heavy quark propagator \eq{eq:quarkpropnonpert} which enters
the Faddeev equation,  and this originates from the fact that in the
generating functional  \eq{eq:genfunc4} we have retained the full
source terms (in contrast with the HQET, where one uses sources for 
the $h$-fields directly).

\section{Summary and conclusions}

In this paper, the Faddeev equation, truncated to include only
two-body interactions for three-quark states in a symmetric
configuration, has been considered. At leading order in the heavy
quark mass expansion  of Coulomb gauge QCD and with the truncation
to include only the nonperturbative two-point functions of the \YM
sector (and neglect all the pure \YM vertices and higher order
functions) the three-quark confining potential has been derived and
a direct connection between the temporal gluon propagator and the
physical string tension has been provided. It was found that, as in
the case of $\bar qq$ systems, the bound state energy increases
linearly with the distance for large separations, and that the
coefficient was $3/2$ times that of the $\bar qq$ system.

Due to the absence of the three-body interactions and the restriction
to a symmetric configuration, no statement about the shape of the
confining string ($\Delta$ or $Y$ configuration) can be made.
From this point of view, a very interesting extension of this work
would be to consider general separations between quarks and to
explicitly include the \YM vertices in the combined system of \DS
and \ST identities, together with the Faddeev equations, and see if
one can extract any information about the shape of the string. Also,
by the inclusion of the \YM vertices one can investigate the charge
screening mechanism, i.e. study how the value of the string tension
modifies.

\begin{acknowledgments}
C.P. has been supported by the Deutscher Akademischer Austausch 
Dienst (DAAD) and partially by the EU-RTN Programme, Contract 
No.MRTN--CT-2006-035482, \lq\lq Flavianet''. P.W. and H.R. have been
supported by the Deutsche Forschungsgemeinschaft (DFG) under
contracts no. DFG-Re856/6-2,3. 
\end{acknowledgments}

\appendix*
\section {Temporal component of the quark-baryon vertex}

In this appendix we present the explicit derivation of the
energy-dependent part of the Faddeev vertex, \eq{eq:gammat}. We start
with \eq{eq:faddeev.simpl} and consider the first of the permutations
of the energy integral:
\be
I=-\int\dk{k_0}\frac{1}
{\left[p_1^0+k_0-m-{\cal I}_r+\imath\e\right]
\left[p_2^0-k_0-m-{\cal I}_r+\imath\e\right]}
\G_{t}(p_1^0+k_0,p_2^0-k_0,p_3^0).
\label{eq:temporal1}
\ee
Using
\be
\frac{1}{\left[z+a+\imath\e\right]
\left[z+b+\imath\e\right]}=\frac{1}{(b-a)}
\left\{\frac{1}{z+a+\imath\e}
-\frac{1}{z+b+\imath\e}\right\}
\label{eq:intid}
\ee
and shifting the integration variables, we find that the integral
$I$, \eq{eq:temporal1}, depends only on the momentum $p_3^0$ (and
implicitly on the bound state energy of  the system $P_0$). 
Explicitly, it reads (using the symmetry of $\G_t$):
\bea
I=-\frac{2}{\left[P_0-p_3^0-2(m+{\cal I}_r)\right]}
\int\dk{k_0}\frac{1}{\left[k_0+P_0-p_3^0-m-
{\cal I}_r+\imath\e\right]}
\G_{t}(P_0-p_3^0+k_0,-k_0,p_3^0).
\eea
Replacing this in the equation \eq{eq:faddeev.simpl}, we find:
\bea
\G_{t}(p_1^0,p_2^0,p_3^0)=-2g^2C_B W_{\si\si}(|\vec r|) 
\sum_{i=1,2,3}
\frac{1}{\left[P_0-p_i^0-2(m+{\cal I}_r)\right]}
\int \dk{k_0}\frac{\G_{t}(P_0-p_i^0+k_0,-k_0,p_i^0)}
{\left[P_0-p_i^0+k_0-m-{\cal  I}_r+\imath\e\right]}.
\label{eq:gammat1}
\eea
The form of the equation \eq{eq:gammat1} suggests that the function
$\G_t$ can be expressed as a symmetric sum
\be
\G_{t}(p_1^0,p_2^0,p_3^0)=f(p_1^0)+f(p_2^0)+f(p_3^0),
\ee
such that the integral equation for $\G_t$ (function of three
variables) is reduced  to an integral equation for the function $f$
(of only one variable). The function $f(p_i^0)$ should be chosen 
such that the integral  on the right hand side of the equation
\eq{eq:gammat1} generates a factor proportional to
$\left[P_0-p_i^0-2(m+{\cal I}_r)\right]$, to cancel the corresponding
factor in the denominator. To examine this possibility, we impose
the following condition:
\be
\frac{1}{\left[P_0-p_i^0-2(m+{\cal I}_r)\right]}
\int \dk{k_0}\frac{f(P_0-p_i^0+k_0)+f(-k_0)+f(p_i^0)}
{\left[P_0-p_i^0+k_0-m-{\cal  I}_r+\imath\e\right]}
=-\frac{\al\; \imath}{P_0-3(m+{\cal I}_r)} f(p_i^0)
\ee
where $\al$ is a (dimensionless) positive constant which remains to
be determined. Rearranging the terms to factorize the function
$f(k_0)$, the above equation can be rewritten as
\be
\int \dk{k_0} f(k_0)
\left[
\frac{1}{k_0-m-{\cal  I}_r+\imath\e}+
\frac{1}{P_0-p_i^0-k_0-m-{\cal  I}_r+\imath\e}
\right]=
(-\imath)\frac{(2\al -1)P_0-2\al p_i^0 
+ (3-4\al) (m+{\cal  I}_r)}{ 2 [P_0-3 (m+{\cal  I}_r)]} f(p_i^0).
\label{eq:gammat2}
\ee
Then the most obvious Ansatz for  the function $f$ is
\be
f(k_0)=\frac{1}{(2\al -1)P_0-2\al k_0 + (3-4\al) 
(m+{\cal  I}_r) +\imath\e}
\ee
such that on the right hand side of the equation \eq{eq:gammat2}
the numerator is cancelled by $f(p_i)$. The next step is to complete
the integration on the on the left hand side, which gives (note that
the $\e$ prescription is chosen such that only the first term in the
bracket survives -- the integration must not give rise to any new
terms containing the energy $p_i^0$):
\bea
\int \dk{k_0} f(k_0) \frac{1}{k_0-m-{\cal  I}_r+\imath\e}
&=&-\frac{\imath}{(2\al -1)P_0 + (3-6\al) (m+{\cal  I}_r)}.
\label{eq:lhs}
\eea
It is then straightforward to compare \eq{eq:gammat2} and \eq{eq:lhs}
and find that the equality is satisfied for $\al=3/2$, leading to the
expression for the vertex $\G_t$ used in the text.

\end{document}